\documentclass[twocolumn]{revtex4}

\usepackage{graphicx}
\usepackage{dcolumn}
\usepackage{amsmath}
\usepackage{amssymb}

\begin{document}

\title{Universal Quantum Computation using Exchange Interactions and
Teleportation of Single-Qubit Operations} 

\author{Lian-Ao Wu and Daniel A. Lidar}

\affiliation{Chemical Physics Theory Group, University of Toronto, 80 
St. George Str., Toronto, Ontario M5S 3H6, Canada}

\begin{abstract}
We show how to construct a universal set of quantum logic gates using
control over exchange interactions and single- and two-spin measurements
only. Single-spin unitary operations are teleported instead of being
executed directly, thus eliminating a major difficulty in the construction
of several of the most promising proposals for solid-state quantum
computation, such as spin-coupled quantum dots, donor-atom nuclear spins
in silicon, and electrons on helium. Contrary to previous proposals
dealing with this difficulty, our scheme requires no encoding
redundancy. We also discuss an application to superconducting phase qubits.
\end{abstract}

\maketitle

Quantum computers (QCs) hold great promise for inherently faster computation
than is possible on their classical counterparts, but so far progress in
building a large-scale QC has been slow. An essential
requirement is that a QC should be capable of performing \textquotedblleft
universal quantum computation\textquotedblright\ (UQC). I.e., it should be
capable of computing, to arbitrary accuracy, any computable function, using a spatially local
and polynomial set of logic gates. One of
the chief obstacles in constructing large scale QCs is the seemingly
innocuous, but in reality very daunting set of requirements that must be met
for universality, according to the standard circuit model \cite{Nielsen:book}: (1) preparation of a fiducial initial state (\textit{initialization}), (2) a set of single and two-qubit unitary transformations
generating the group of all unitary transformations on the Hilbert space of
the QC (\textit{computation}), and (3) single-qubit measurements (\textit{read-out}). Since initialization can often be performed through
measurements, requirements (1) and (3) do not necessarily imply different
experimental procedures and contraints. Until recently it was thought that
computation is irreducible to measurements, so that requirement (2), a set
of unitary transformations, would appear to be an essential component of UQC. However,
unitary transformations are sometimes very challenging to perform. Two
important examples are the exceedingly small photon-photon interaction that
was thought to preclude linear optics QCs, and the difficult to execute
single-spin gates in certain solid state QC proposals, such as quantum dots 
\cite{Loss:98,Levy:01a} and donor atom nuclear spins in silicon \cite{Kane:98,Mozyrsky:01}. The problem with single-spin unitary gates is that
they impose difficult demands on $g$-factor engineering of heterostructure
materials, and require strong and inhomogeneous magnetic fields or microwave
manipulations of spins, that are often slow and may cause device heating. In
the case of exchange Hamiltonians, a possible solution was recently proposed in terms of
qubits that are encoded into the states of two or more spins, whence
the exchange interaction alone is sufficient to construct a set of universal
gates \cite{Bacon:99a,Kempe:00,DiVincenzo:00a,Bacon:Sydney,Levy:01,Benjamin:01,LidarWu:01,WuLidar:01,Kempe:01,Kempe:01a,WuLidar:01a,WuLidar:02,Vala:02,Skinner:02}
(the \textquotedblleft encoded universality\textquotedblright\ approach). In
the linear optics case, it was shown that photon-photon interactions can be
induced indirectly via \emph{gate teleportation} \cite{Knill:00}. This idea
has its origins in earlier work on fault-tolerant constructions for quantum
gates \cite{Shor:96,Preskill:97a,Gottesman:99b} (generalized in \cite{Zhou:00}) and
stochastic programmable quantum gates \cite{Nielsen:97a,Vidal:02}. The
same work inspired more recent results showing that, in fact, measurements
and state preparation \emph{alone} suffice for UQC \cite{Nielsen:01,Fenner:01,Leung:01a,Raussendorf:01}.

Experimentally, a minimalistic approach to constructing a QC seems
appealing. In this sense, retaining only the absolutely essential
ingredients needed to construct a universal QC may be an important
simplification. Since read-out is necessary, \emph{measurements are
inevitable}. Here we propose a minimalistic approach for universal quantum
computation that is particularly well suited to the important class of
spin-based QC proposals governed by exchange interactions \cite{Loss:98,Levy:01a,Kane:98,Vrijen:00,Mozyrsky:01,Imamoglu:99}, and other
proposals governed by \emph{effective} exchange interactions \cite{Cirac:95,Platzman:99,Brown:01}. In particular, we show that \emph{UQC can
be performed using only single- and two-qubit measurements and controlled
exchange interactions, via gate teleportation}. In our approach, which offers
a new perspective on the requirements for UQC, the need to perform the
aforementioned difficult single-spin unitary operations is obviated, and
replaced by measurements, which are anyhow necessary. The tradeoff is that
the implementation of gates becomes probabilistic (as in all
gate-teleportation based approaches), but this probability can be boosted
arbitrarily close to 1 exponentially fast in the number of measurements.

We begin our discussion with a relatively simple example of the utility
of measurement-aided UQC. This example is not in the exchange-interaction
category, but both serves to illustrate some of the more complex ideas
needed below, and solves a problem of relevance to an important
solid-state QC proposal. The proposal we have in mind is that using d-wave grain boundary
(dGB) phase qubits 
\cite{Zagoskin:99,Blais:00}. The system Hamiltonian is: 
\begin{equation}
H_{S}=H_{X}+H_{Z}+H_{ZZ},
\end{equation}
where 
$H_{X} =\sum_{i}\Delta _{i}X_{i}$ describes phase
tunneling, $H_{Z} =\sum_{i}b_{i}Z_i$ is a bias, and
$H_{ZZ} =\sum_{i,j} J_{ij}Z_i Z_j$
represents Josephson coupling of qubits; $X_{i},Y_{i},Z_{i}$ denote the Pauli matrices $\sigma
^{x},\sigma ^{y},\sigma ^{z}$ acting on the $i^{\mathrm{th}}$ qubit. It turns out that in this system
{\em only one} of the terms $H_{X},H_{Z},H_{ZZ}$ can be on at any
given time \cite{Zagoskin:99,Blais:00}. Moreover, turning on the bias or
Josephson coupling is the only way to control the value of the tunneling
matrix element. In the idle state $\Delta_i$ is non-zero and the qubit undergoes coherent tunneling.
In the dGB proposal it is important to 
reduce the constraints on fabrication by removing the
possibility of applying bias $b_{i}$ on individual qubits
\cite{LidarWuBlais:02}. This bias 
requires, e.g., the possibility of applying a local magnetic field on
each qubit, and is experimentally very challenging to realize. 
The effective system Hamiltonian that we consider is
therefore: $H_{S}' = H_X + H_{ZZ}$, with continuous control over $J_{ij}$. In
\cite{LidarWuBlais:02} it was shown how UQC can be performed given
this Hamiltonian, by encoding a logical qubit into two physical
qubits, and using sequences of recoupling pulses. Here we show instead how to
implement $Z_{i}$ using measurements, which together with $H_{S}'$ is
sufficient for UQC. Suppose we start from an unknown state of qubit 1: $\left| \psi
\right\rangle =a$  $\left| 0\right\rangle +b$ $\left| 1\right\rangle
$. By cooling in the idle state (only $H_{X}$ on) we can prepare an ancilla qubit 2 in the state $(\left| 0\right\rangle +$ $\left|
1\right\rangle )/\sqrt{2}$. Then the total state is: $a\left| 00\right\rangle +b\left| 10\right\rangle +a\left| 01\right\rangle
+b\left| 11\right\rangle$. Letting the Josephson-gate $e^{-i\phi Z_{1}Z_{2}/2}$ act on this
state, we obtain
\begin{eqnarray*}
&&a\left| 00\right\rangle +e^{i\phi }b\left| 10\right\rangle +e^{i\phi
}a\left| 01\right\rangle +b\left| 11\right\rangle  \\
&\propto &e^{-i\phi Z_{1}/2}\left| \psi \right\rangle \left|
0\right\rangle +e^{i\phi Z_{1}/2}\left| \psi \right\rangle \left|
1\right\rangle 
\end{eqnarray*}
We then measure $Z_{2}$. If we find $0$ (with
probablity $1/2$) then the state has collapsed to $e^{-i\phi
Z_{1}/2}\left| \psi \right\rangle \left| 0 \right\rangle$, which is the required
operation on qubit 1. If we find $-1$
then the state is $e^{i\phi Z_{1}/2}\left| \psi \right\rangle  \left| 1 \right\rangle$, which is
an erred state. To correct it we apply the pulse $e^{-i\phi Z_{1}Z_{2}}$, which takes the erred state
to the correct state $-e^{-i\phi Z_{1}/2}\left| \psi \right\rangle
 \left| 1 \right\rangle $. We then
re-initialize the ancilla qubit. This method for implementing $Z_{i}$ succeeds with certainty
after one measurement, possibly requiring (with probability $1/2$) one
correction step.

We now turn to QC-proposals based on exchange interactions \cite{Loss:98,Levy:01a,Kane:98,Vrijen:00,Mozyrsky:01,Imamoglu:99,Cirac:95,Platzman:99,Brown:01}.
In these systems, that are some of the more promising candidates for scalable QC, 
the qubit-qubit interaction can be written as an axially symmetric
exchange interaction of the form:
\begin{equation}
H_{ij}^{\mathrm{ex}}(t)=J_{ij}^{\perp}(t)(X_{i}X_{j}+Y_{i}Y_{j})+J_{ij}^{z}(t)Z_{i}Z_{j},  \label{eq:Hex}
\end{equation}
where $J_{ij}^{\alpha }(t)\ $ ($\alpha =\perp ,z$) are controllable
coupling constants.
The XY
(XXZ) model is the case when $J_{ij}^{z}=0$ ($\neq 0$). The Heisenberg
interaction is the case when $J_{ij}^{z}(t)=J_{ij}^{\perp }(t)$. See \cite{LidarWu:01} for a classification of various QC models by the type of
exchange interaction. In agreement with the QC proposals \cite{Loss:98,Levy:01a,Kane:98,Vrijen:00,Mozyrsky:01,Imamoglu:99,Cirac:95,Platzman:99,Brown:01}, we assume here that $J_{ij}^{\perp }(t)$ is competely controllable and
allow that the ratio between $J_{ij}^{\perp }(t)$ and $J_{ij}^{z}(t)$ may
not be controllable. The method we present here works equally well for all
three types of exchange interactions, thus unifying all exchange-based
proposals under a single universality framework. Since all terms in $H_{
\mathrm{ex}}(t)$ commute it is simple to show that it generates a unitary
two-qubit evolution operator of the form $U_{ij}(\varphi ^{\perp },\varphi
^{z})=\exp [-i\int^{t}dt^{\prime }H_{ij}^{\mathrm{ex}}(t^{\prime })]=$ 
\begin{equation}
\left( 
\begin{array}{cccc}
e^{-i\varphi ^{z}} &  &  &  \\ 
& e^{i\varphi ^{z}}\cos 2\varphi ^{\perp } & -ie^{i\varphi ^{z}}\sin
2\varphi ^{\perp } &  \\ 
& -ie^{i\varphi ^{z}}\sin 2\varphi ^{\perp } & e^{i\varphi ^{z}}\cos
2\varphi ^{\perp } &  \\ 
&  &  & e^{-i\varphi ^{z}}
\end{array}
\right)  \label{eq:1}
\end{equation}
(we use units where $\hbar =1$), where $\varphi ^{\alpha
}=\int^{t}dt^{\prime }J^{\alpha }(t^{\prime })$, and we have suppressed the qubit indices for clarity.
In preparation of our main result, we first prove:

\noindent \textit{Proposition}.
The set $\mathcal{G}=\{U_{ij}(\varphi ^{\perp },\varphi ^{z}), R_{j\beta
}\equiv \exp (i\frac{\pi }{4}\sigma _{j}^{\beta })\}$ ($\beta =x,z$) is
universal for quantum computation.

\emph{Proof}: A set of continuous one-qubit unitary gates and any two-body
Hamiltonian entangling qubits are universal for quantum computation
\cite{Dodd:01}. The exchange Hamiltonian $H^{\rm ex}_{ij}$ clearly can
generate entanglement, so it suffices to show that
we can generate all single-qubit transformations using $\mathcal{G}$. Two of
the Pauli matrices are given simply by $\sigma _{j}^{\beta }=-iR_{j\beta
}^{2}$. Now, let $C_{A}^{\theta }\circ \exp (i\varphi B)\equiv \exp
(-i\theta A)\exp (i\varphi B)\exp (+i\theta A)$; two useful identities for
anticommuting $A,B$ with $A^{2}=I$ (the identity) are \cite{WuLidar:01a}: 
\[
C_{A}^{\pi /2}\circ e^{-i\varphi B}=e^{i\varphi B},\quad C_{A}^{\pi /4}\circ
e^{-i\varphi B}=e^{\varphi AB}.
\]
Using this, we first generate $e^{-i\varphi X_{1}X_{2}}=U_{12}(\varphi
/2,\varphi ^{z})C_{X_{1}}^{\pi /2}\circ U_{12}(\varphi /2,\varphi ^{z})$,
which takes six elementary steps (where an elementary step is defined
as one of the operations $U_{ij}(\varphi ^{\perp},\varphi ^{z}), R_{j\beta}$). Second, as we show below,
our gate teleportation
procedure can prepare $R_{j\beta }^{\dagger}$ just as efficiently as
$R_{j\beta }^{\dagger}$ (also note that $R_{j\beta
}^{\dagger}=-(R_{j\beta })^{3}$), so that with two additional steps we have $e^{-i\varphi
Y_{1}X_{2}}=C_{Z_{1}}^{-\pi /4}\circ $ $e^{-i\varphi X_{1}X_{2}}$. Finally,
with a total of $8+6+8=22$ elementary steps we have $e^{-i\varphi
Z_{1}}=C_{Y_{1}X_{2}}^{\pi /4}\circ e^{-i\varphi X_{1}X_{2}}$, where $
\varphi $ is arbitrary. Similarly, we can generate $e^{-i\varphi Y_{1}}$ in
22 steps using $C_{X_{1}}^{\pi /4}$ instead of $C_{Z_{1}}^{-\pi /4}$. Using
a standard Euler angle construction we can generate arbitrary single-qubit
operations by composing $e^{-i\varphi Z_{1}}$ and $e^{-i\varphi Y_{1}}$ \cite{Nielsen:book}.

\begin{figure}[tbp]
\includegraphics[height=15cm,angle=270]{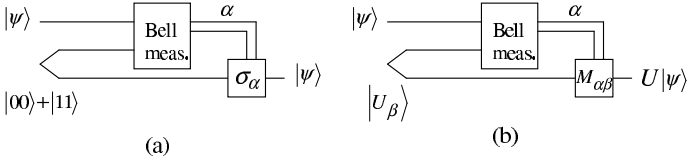}
\vspace{-8.5cm}
\caption{Teleporation \protect\cite{Bennett:93} is a method for transmitting
an unknown quantum state $|\protect\psi \rangle $ with the help of prior
entanglement and classical communication. A state teleportation circuit is
shown in (a), where time proceeds from left to right, and $<$ denotes the
entangled (Bell) state $\frac{1}{\protect\sqrt{2}} (|00\rangle
_{23}+|11\rangle _{23})$. Alice has $|\protect\psi \rangle _{1}$ and qubit $
2 $ from the Bell state. Bob has qubit $3$ from the Bell state. Alice
measures $|\protect\psi \rangle _{1}$ and qubit $2$ in the Bell basis,
obtaining one of $4$ possible outcomes labeled $\protect\alpha $. She
communicates her result to Bob (double wires), who applies $\protect\sigma 
^{ \protect\alpha }$ to his qubit, where $\protect\sigma ^{\protect\alpha }$
are the four Pauli matrices $I,\protect\sigma ^{x},\protect\sigma ^{y}, 
\protect\sigma ^{z}$. Bob then has $|\protect\psi \rangle _{3}$. A gate
teleportation circuit is shown in (b), following \protect\cite{Nielsen:01}.
To teleport the single-qubit operation $U$, the state $|U_{\protect\beta 
}\rangle \equiv (I\otimes U\protect\sigma ^{\protect\beta })\frac{1}{ 
\protect\sqrt{2}}(|00\rangle +|11\rangle )$ is prepared offline, by first
preparing the state $|00\rangle $ and then measuring in the orthonormal
basis of states $|U_{\protect\beta }\rangle $. Alice and Bob now repeat the
state teleportation protocol. With probability $1/4$ Alice finds $\protect
\alpha = \protect\beta $, in which case Bob now has $U|\protect\psi \rangle
_{3}$. With probability $3/4$ she finds $\protect\alpha \neq \protect\beta $
and Bob needs to apply a correction $M_{\protect\alpha \protect\beta }=U 
\protect\sigma ^{\protect\beta }\protect\sigma ^{\protect\alpha }U^{\dag }$
in order to end up with $U|\protect\psi \rangle _{3}$. This is done by
teleporting $M_{\protect\alpha \protect\beta }$, i.e., the procedure is
repeated recursively. It succeeds on average after $4$ trails.}
\label{fig1}
\end{figure}

\begin{figure}
\centering
\includegraphics[height=15cm,angle=270]{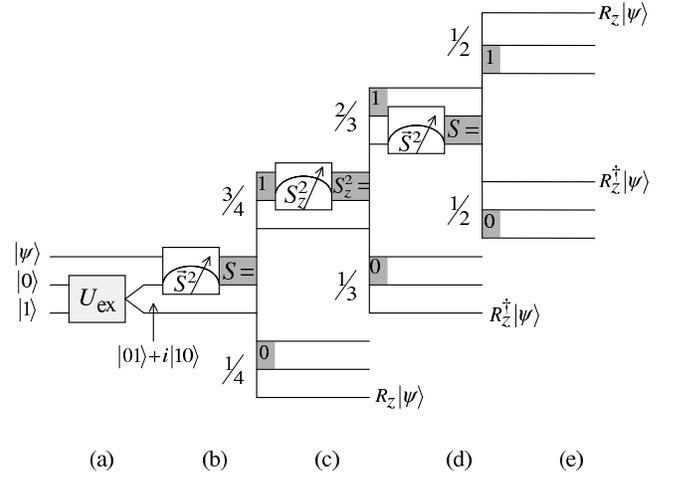}
\vspace{-5cm}
\caption{Gate teleportation of single-qubit operation
$R_{z}$. Initially Alice
has $|\protect\psi \rangle _{1}$ and $ \left| 0 \right\rangle$. Bob has $ \left| 1 \right\rangle$. Time proceeds from left to right. Starting from the $3$-qubit state $|\protect\psi \rangle
|01\rangle $, the task is to obtain $R_{z}|\protect\psi \rangle $. The
protocol shown succeeds with probability $1/2$. When it fails the operation $
R_{z}^{\dagger }$ is applied instead. Fractions give
the probability of a branch; $0$ and $1$ in a gray box are possible
measurement outcomes of the observable in the preceeding gray box. See
text for full details.}
\label{fig2}
\end{figure}

It is important to note that optimization of the number of steps given in the proof above 
may be possible. We now show that the single qubit gates $R_{j\beta }$ can be
implemented using cooling, weak spin measurements, and evolution under exchange
Hamiltonians of the Heisenberg, XY, or XXZ type. Our method is inspired by the gate teleportation
idea
\cite{Nielsen:97a,Shor:96,Preskill:97a,Gottesman:99b,Zhou:00,Vidal:02,Knill:00,Nielsen:01,Fenner:01,Leung:01a}, which we briefly review, along with state teleportation
\cite{Bennett:93}, in Fig.~\ref{fig1}.
We proceed in two cycles. In Cycle (i), consider a spin (our
\textquotedblleft data qubit\textquotedblright) in an
unknown state $\left\vert \psi \right\rangle =a\left\vert 0\right\rangle
+b\left\vert 1\right\rangle $, and two additional (\textquotedblleft
ancilla\textquotedblright ) spins, as shown in Fig.~\ref{fig2}. Our task is
to apply the one-qubit operation $R_{\beta }$ to the data qubit. As in gate
teleporation, we require an entangled pair of ancilla spins. However, it
turns out that rather than one of the Bell states we need an entangled state
that has a phase of $i$ between its components. To obtain this state, we
first turn on the exchange interaction $H_{23}^{\mathrm{ex}}$ between the
ancilla spins such that $J^{\perp }>0$. The eigenvalues (eigenstates) are $
\{-2J^{\perp }-J^{z},2J^{\perp }-J^{z},J^{z},J^{z}\}$ ($\left\vert
S\right\rangle ,\left\vert T_{0}\right\rangle ,\left\vert 00\right\rangle
,\left\vert 11\right\rangle $) where $\left\vert S\right\rangle =\frac{1}{
\sqrt{2}}(\left\vert 01\right\rangle -\left\vert 10\right\rangle )$, $
\left\vert T_{0}\right\rangle =\frac{1}{\sqrt{2}}(\left\vert 01\right\rangle
+\left\vert 10\right\rangle )$ are the singlet and one of the triplet
states. Provided $J^{\perp }>-J^{z}$ [which is the case for all QC proposals
of interest, in which either $\mathrm{sign}(J^{\perp })=\mathrm{sign}(J^{z})$, or $J^{z}=0$] and we cool the system significantly below $-2J^{\perp
}-J^{z}$, the resulting ground state is $\left\vert S\right\rangle $. We
then perform a single-spin measurement of the observable $\sigma _{j}^{z}$
on one or both of the ancillas, which will yield either $\left\vert 01\right\rangle $ or $
\left\vert 10\right\rangle $. For definiteness assume the outcome was $
\left\vert 01\right\rangle $. We then immediately apply a $\pi /8$ exchange
pulse to the ancilla spins [Fig.~ \ref{fig2}(a)]: $U(\pi /8,\varphi
_{0}^{z})\left\vert 10\right\rangle =\frac{e^{i\varphi ^{z}}}{\sqrt{2}}
(|01\rangle -i|10\rangle )$ [as follows from Eq.~(\ref{eq:1})]. The total
state of the three spins then reads (neglecting an overall phase $
e^{i\varphi ^{z}}$):
\begin{eqnarray}
\left\vert \psi \right\rangle _{1}U_{23}(\pi /8,\varphi _{0}^{z})\left\vert
10\right\rangle _{23} =\frac{1}{\sqrt{2}}(a\left\vert 001\right\rangle
-ib\left\vert 110\right\rangle )  \nonumber \\
+\frac{1}{2}r\left\vert T_{0}\right\rangle _{12}R_{3z}^{\dagger }\left\vert
\psi \right\rangle _{3}-\frac{1}{2}r^{\ast }\left\vert S\right\rangle
_{12}R_{3z}\left\vert \psi \right\rangle _{3}  \label{eq:2}
\end{eqnarray}
where $r=\exp (-i\pi /4)$ and the subscripts denote the spin index.

At this point Alice makes a weak measurement
of her spins [Fig.~\ref{fig2}(b)]. Let $\overrightarrow{S}_{ij}=\frac{1}{2}(\vec{\sigma}
_{i}+\vec{\sigma}_{j})$ be the total spin of qubits $i,j$; Alice measures $ 
\vec{S}_{12}^{2}$, with eigenvalues $S(S+1)$. Since only for the singlet
state $\left\vert S\right\rangle _{12}$ do we have $S(S+1)=0$, it follows
that if the measurement yields $0$, then the state has collapsed to $
\left\vert S\right\rangle _{12}R_{3z}\left\vert \psi \right\rangle _{3}$. In
this case, which occurs with probability $1/4$, Bob has $R_{3z}\left\vert
\psi \right\rangle _{3}$, and we are done [Fig.~\ref{fig2}(c), bottom]. If,
on the other hand, Alice finds $S=1$, then the normalized post-measurement
state is 
\begin{equation}
\frac{1}{\sqrt{3}}[r\left\vert T_{0}\right\rangle _{12}R_{3z}^{\dagger
}\left\vert \psi \right\rangle _{3}+a\sqrt{2}(\left\vert 001\right\rangle
-ib\left\vert 110\right\rangle )].  \label{eq:3}
\end{equation}
Similar to the gate teleportation protocol \cite{Nielsen:01,Fenner:01,Leung:01a}
shown in Fig.~\ref{fig1}(b),
Alice and Bob now need to engage in a series of correction steps. In the
next step Alice measures $S_{z}^{2}=\frac{1}{4}(\sigma _{1}^{z}+\sigma
_{2}^{z})^{2}=\frac{1}{2}(I+\sigma _{1}^{z}\sigma _{2}^{z})$ [Fig.~\ref{fig2}
(c), top]. Measurement of the observable $\sigma _{1}^{z}\sigma _{2}^{z}$ is
discussed in \cite{Nielsen:book}. If Alice finds $S_{z}^{2}=0$ then with
probability $1/3$ the state collapses to $\left\vert T_{0}\right\rangle
_{12}R_{3z}^{\dagger }\left\vert \psi \right\rangle _{3}$ and Bob ends up
with the opposite of the desired operation, namely $R_{z}^{\dagger
}\left\vert \psi \right\rangle $ [Fig.~\ref{fig2}(d), bottom]. We describe
the required corrective action below, in Cycle (ii). If Alice finds $S_{z}^{2}=1$, then the
state is: 
\[
a\left\vert 001\right\rangle -ib\left\vert 110\right\rangle =\frac{1}{\sqrt{
2 }}(r^{\ast }R_{1z}^{\dagger }\left\vert \psi \right\rangle _{1}\left\vert
S\right\rangle _{23}+rR_{1z}\left\vert \psi \right\rangle _{1}\left\vert
T_{0}\right\rangle _{23}).
\]
Bob now measures $\vec{S}_{23}^{2}$. If he finds $S=0$
then the state has collapsed to $R_{1z}^{\dagger }\left\vert \psi
\right\rangle _{1}\left\vert S\right\rangle _{23}$, while if $S=1$ then the
outcome is $R_{1z}\left\vert \psi \right\rangle _{1}\left\vert
T_{0}\right\rangle _{23}$, equiprobably. In the latter case
Alice ends up with the desired operation [Fig.~\ref{fig2}(e)].

In a similar manner one can generate $R_{x}$ or $R_{x}^{\dagger }$ acting on
an arbitrary qubit state $\left\vert \psi \right\rangle $. Let $\left\vert
\pm \right\rangle $ denote the $\pm 1$ eigenstates of the Pauli operator $
\sigma ^{x}$. As in the $R_{z}$ case above, first prepare a singlet state $
\left\vert S\right\rangle =\frac{1}{\sqrt{2}}(\left\vert -+\right\rangle
-\left\vert +-\right\rangle )$ on the ancilla spins $2,3$ by cooling. Then
perform a single-spin measurement of the observable $\sigma _{j}^{x}$ on
each ancilla, which will yield either $\left\vert +-\right\rangle $ or $
\left\vert -+\right\rangle $. For definiteness assume the outcome was $
\left\vert +-\right\rangle _{23}$. Observing that in the $\{|+-\rangle
,|-+\rangle \}$ subspace, $H_{ij}^{\mathrm{ex}}=-J_{ij}^{\perp
}I+(J_{ij}^{\perp }+J_{ij}^{z})\tilde{X}$, where $\tilde{X}:|+-\rangle
\leftrightarrow |-+\rangle $, it follows that $U(\pi /4-\varphi
_{0}^{z},\varphi _{0}^{z})\left\vert +-\right\rangle =\frac{e^{-i\varphi
^{\perp }}}{\sqrt{2}}(|+-\rangle -i|-+\rangle )$, so that we have a means of
generating an entangled initial state. The unknown state $\left\vert \psi
\right\rangle _{1}$ of the data qubit can be expressed as $\left\vert \psi
\right\rangle =a_{x}\left\vert +\right\rangle +b_{x}\left\vert
-\right\rangle $, where $a_{x}=(a+b)/\sqrt{2}$ and $b_{x}=(a-b)/\sqrt{2}.$
Then (neglecting the overall phase $e^{-i\varphi ^{\perp }}$): 
\begin{eqnarray*}
\left\vert \psi \right\rangle _{1}U_{23}(\pi /4-\varphi _{0}^{z},\varphi
_{0}^{z})\left\vert +-\right\rangle _{23} =\frac{1}{2}r^{\ast }\left\vert
S\right\rangle _{12}R_{3x}\left\vert \psi \right\rangle _{3}+ \\
\frac{1}{2}r\left\vert T_{0}^{x}\right\rangle _{12}R_{3x}^{\dagger
}\left\vert \psi \right\rangle _{3}+\frac{1}{\sqrt{2}}(a_{x}\left\vert
++-\right\rangle -ib_{x}\left\vert --+\right\rangle )
\end{eqnarray*}
where $\left\vert T_{0}^{x}\right\rangle =\frac{1}{\sqrt{2}}(\left\vert
+-\right\rangle +\left\vert -+\right\rangle )$ is a triplet state, a zero
eigenstate of the observable $\sigma _{1}^{x}+\sigma _{2}^{x}$. The gate
teleportation procedure is now repeated to yield $R_{x}$ or $R_{x}^{\dagger
}.$ First, Alice measures the total spin $\vec{S}_{12}^{2}$. If she find $S=0
$ (with probability $1/4$) Bob has spin $3$ in the desired state $
R_{3x}\left\vert \psi \right\rangle _{3}$. If she finds $S=1$ then she
proceeds \ to measure the total length of the $x$\textit{\ }component $
S_{x}^{2}=\frac{1}{4}(\sigma _{1}^{x}+\sigma _{2}^{x})^{2}$, yielding,
provided she finds $S_{x}^{2}=0$, the state $\left\vert
T_{0}^{x}\right\rangle _{12}R_{3x}^{\dagger }\left\vert \psi \right\rangle
_{3}$ with probability $1/3$. If, on the other hand, she finds $S_{x}^{2}=1$, i.e., the state is $a_{x}\left\vert ++-\right\rangle -ib_{x}\left\vert
--+\right\rangle $, then by letting Bob measure $\vec{S}_{23}^{2}$, the
states $R_{1x}^{\dagger }\left\vert \psi \right\rangle _{1}\left\vert
S\right\rangle _{23}$ or $R_{1x}\left\vert \psi \right\rangle _{1}\left\vert
T_{0}^{x}\right\rangle _{23}$ are obtained, with equal probabilities.

Fig.~\ref{fig2} summarizes the protocol we have described thus far. The
overall effect is to transform the input state $\left\vert \psi
\right\rangle $ to either the output state $R_{\beta }\left\vert \psi
\right\rangle $ or $R_{\beta }^{\dagger }\left\vert \psi\right\rangle $,
equiprobably.

We have now arrived at Cycle (ii), in which we must fix the erred state $
R_{j\beta }^{\dagger }\left\vert \psi\right\rangle _{j}$ ($j=1$ or
$3$). To do so we
essentially repeat the procedure shown in Fig.~\ref{fig2}. We
explicitly discuss one example; all other cases are
similar. Suppose that we obtain the erred state $R_{1z}^{\dagger }\left\vert
\psi\right\rangle _{1}\left\vert S\right\rangle _{23}$
[Fig.~\ref{fig2}(e)]. It can be rewritten as
\begin{eqnarray*}
r R_{1z}^{\dagger }\left\vert
\psi\right\rangle _{1}\left\vert S\right\rangle _{23} = -\frac{i}{\sqrt{2}}(a\left\vert 001\right\rangle
-ib\left\vert 110\right\rangle ) \\
-\frac{1}{2}r\left\vert S\right\rangle_{12}R_{3z}^{\dagger }\left\vert
\psi\right\rangle_{3}+\frac{1}{2}r^*\left\vert
T_{0}\right\rangle_{12}R_{3z}\left\vert \psi\right\rangle _{3},
\end{eqnarray*}
which up to unimportant phases is identical to Eq.~(\ref{eq:2}), except that the position of $
R_{3z}^{\dagger }$ and $R_{3z}$ has flipped. Correspondingly flipping the
decision pathway in Fig.~\ref{fig2} will therefore lead to the correct
action $R_{\beta }\left\vert \psi\right\rangle $ with probability $1/2$,
while the overall probability of obtaining the faulty outcome $R_{\beta
}^{\dagger }\left\vert \psi\right\rangle $ after the second cycle of
measurements is $1/4.$ Clearly, after $n$ measurement cycles as shown in
Fig.~\ref{fig2}, the probability for the correct outcome is
$1-2^{-n}$. The expected number of measurements per
cycle is $1 \frac{1}{4} + 3 \frac{3}{4}\frac{2}{3}\frac{1}{2} =
1$, and the expected number of measurement cycles needed is $\sum_{n=1}^{\infty
}n2^{-n}=2$.

We note that in the case of the erred state $R_{jz}^{\dagger }\left\vert
\psi \right\rangle _{j}$ ($j=1$ or $3$) there is an alternative that is
potentially simpler than repeating the measurement scheme of Fig.~\ref{fig2}. Provided the exchange Hamiltonian is of the XY type, or of the XXZ type
with a tunable $J^{z}$ exchange parameter, one can simply apply the correction operator $U_{j2}(\frac{\pi }{2}
,0)=Z_{j}Z_{2}$ to $R_{jz}^{\dagger }\left\vert \psi \right\rangle _{j}$,
yielding $R_{jz}\left\vert \psi \right\rangle _{j}$ as required. Finally, we
note that Nielsen \cite{Nielsen:01} has discussed the conditions for making
a gate teleportation procedure of the type we have proposed here, fault
tolerant.

To conclude, we have proposed a gate-teleportation method for
universal quantum computation that is uniformly applicable to Heisenberg, XY
and XXZ-type exchange interaction-based quantum computer (QC) proposals.
Such exchange interactions characterize almost all solid-state QC
proposals, as well as several quantum optics based proposals \cite{LidarWu:01}. In a number of these QC proposals, e.g., quantum dots \cite{Loss:98}, exchange interactions are significantly easier to control than
single-qubit operations \cite{DiVincenzo:00a,LidarWu:01}. Therefore it is
advantageous to replace, where possible, single-qubit operations by
measurements. Moreover, spin measurements are necessary for state read-out,
both at the end of a computation and at intermediate stages during an
error-correction prodecure, and often play an important role in
initial-state preparation. Our method combines measurements of single- and
two-spin observables, and a tunable exchange interaction. In a similar
spirit we have shown how to replace with
measurements certain difficult single-qubit operations
in a QC-proposal involving superconducting phase qubits. We hope that the
flexibility offered by this approach will provide a useful alternative route
towards the realization of universal quantum computation.

{\it Acknowledgments}.--- The present study was sponsored by the DARPA-QuIST 
program (managed by AFOSR under agreement No. F49620-01-1-0468) and by DWave 
Systems, Inc.


\begin{references}

\bibitem{Nielsen:book}
{M.A. Nielsen and I.L. Chuang}, {\it {Quantum Computation and Quantum
  Information}\/} ({Cambridge University Press}, Cambridge, UK, 2000).

\bibitem{Loss:98}
{D. Loss, D.P. DiVincenzo}, {\it Phys. Rev. A\/} {\bf 57}, 120 (1998).

\bibitem{Levy:01a}
{J. Levy}, {\it Phys. Rev. A\/} {\bf 64}, 052306 (2001).

\bibitem{Kane:98}
{B.E. Kane}, {\it Nature\/} {\bf 393}, 133 (1998).

\bibitem{Mozyrsky:01}
{D. Mozyrsky, V. Privman, M.L. Glasser}, {\it Phys. Rev. Lett.\/} {\bf 86},
  5112 (2001).

\bibitem{Bacon:99a}
{D. Bacon, J. Kempe, D.A. Lidar, K.B. Whaley}, {\it Phys. Rev. Lett.\/} {\bf
  85}, 1758 (2000).

\bibitem{Kempe:00}
{J. Kempe, D. Bacon, D.A. Lidar, K.B. Whaley}, {\it Phys. Rev. A\/} {\bf
  63}, 042307 (2001).

\bibitem{DiVincenzo:00a}
{D.P. DiVincenzo, D. Bacon, J. Kempe, G. Burkard, K.B. Whaley}, {\it
  Nature\/} {\bf 408}, 339 (2000).

\bibitem{Bacon:Sydney}
{D. Bacon, J. Kempe, D.P. DiVincenzo, D.A. Lidar, K.B. Whaley}, {\it
  Proceedings of the 1st International Conference on Experimental
  Implementations of Quantum Computation, Sydney, Australia\/}, R.~Clark, ed.
  (Rinton, Princeton, NJ, 2001), p. 257.

\bibitem{Levy:01}
{J. Levy}, Eprint quant-ph/0101057.

\bibitem{Benjamin:01}
{S.C. Benjamin}, {\it Phys. Rev. A\/} {\bf 64}, 054303 (2001).

\bibitem{LidarWu:01}
{D.A. Lidar, L.-A. Wu}, {\it Phys. Rev. Lett.\/} {\bf 88}, 017905 (2002).

\bibitem{WuLidar:01}
{L.-A. Wu, D.A. Lidar}, {\it Phys. Rev. A\/} {\bf 65}, 042318 (2002).

\bibitem{Kempe:01}
{J. Kempe, D. Bacon, D.P. DiVincenzo, K.B. Whaley}, {\it Quant. Inf.
  Comp.\/} {\bf 1}, 33 (2001).

\bibitem{Kempe:01a}
{J. Kempe, K.B. Whaley}, {\it Phys. Rev. A\/} {\bf 65}, 052330 (2001).

\bibitem{WuLidar:01a}
{L.-A. Wu, D.A. Lidar}, {\it J. Math. Phys.\/}, in press. Eprint quant-ph/0109078.

\bibitem{WuLidar:02}
{L.-A. Wu, D.A. Lidar}, Eprint quant-ph/0202135.

\bibitem{Vala:02}
{J. Vala, K.B. Whaley}, Eprint quant-ph/0204016.

\bibitem{Skinner:02}
{A.J. Skinner, M.E. Davenport, B.E. Kane}, Eprint quant-ph/0206159.

\bibitem{Knill:00}
{E. Knill, R. Laflamme, G.J. Milburn}, {\it Nature\/} {\bf 409}, 46 (2001).

\bibitem{Shor:96}
{P.W. Shor}, {\it {Proceedings of the 37th Symposium on Foundations of
  Computing}\/} ({IEEE Computer Society Press}, {Los Alamitos, CA}, 1996),
  p.~56.

\bibitem{Preskill:97a}
{J. Preskill}, {\it Proc. Roy. Soc. London Ser. A} {\bf 454}, 385 (1998).

\bibitem{Gottesman:99b}
{D. Gottesman, I.L. Chuang}, {\it Nature\/} {\bf 402}, 390 (1999).

\bibitem{Zhou:00}
{X. Zhou, D.W. Leung, I.L. Chuang}, {\it Phys. Rev. A\/} {\bf 62}, 052316
  (2000).

\bibitem{Nielsen:97a}
{M.A. Nielsen, I.L. Chuang}, {\it Phys. Rev. Lett.\/} {\bf 79}, 321 (1997).

\bibitem{Vidal:02}
{G. Vidal, L. Masanes, J.I. Cirac}, {\it Phys. Rev. Lett.\/} {\bf 88}, 047905
  (2002).

\bibitem{Nielsen:01}
{M.A. Nielsen}, Eprint quant-ph/0108020.

\bibitem{Fenner:01}
{S.A. Fenner, Y. Zhang}, Eprint quant-ph/0111077.

\bibitem{Leung:01a}
{D.W. Leung}, Eprint quant-ph/0111122.

\bibitem{Raussendorf:01}
{R. Raussendorf, H.J. Briegel}, {\it Phys. Rev. Lett.\/} {\bf 86}, 5188
  (2001).

\bibitem{Vrijen:00}
{R. Vrijen {\it et al.}}, {\it Phys. Rev. A\/} {\bf 62}, 012306 (2000).

\bibitem{Imamoglu:99}
{A. Imamo$\bar{\rm g}$lu {\it et al.}}, {\it Phys. Rev. Lett.\/} {\bf 83}, 4204 (1999).

\bibitem{Cirac:95}
{J.I. Cirac, P. Zoller}, {\it Phys. Rev. Lett.\/} {\bf 74}, 4091 (1995).

\bibitem{Platzman:99}
{P.M. Platzman, M.I. Dykman}, {\it Science\/} {\bf 284}, 1967 (1999).

\bibitem{Brown:01}
{K.R. Brown, D.A. Lidar, K.B. Whaley}, {\it Phys. Rev. A\/} {\bf 65}, 012307
  (2002).

\bibitem{Zagoskin:99}
{A.M. Zagoskin}, Eprint cond-mat/9903170.

\bibitem{Blais:00}
{A. Blais, A.M. Zagoskin}, {\it Phys. Rev. A\/} {\bf 61}, 042308 (2000).

\bibitem{LidarWuBlais:02}
{D.A. Lidar, L.-A. Wu, A. Blais}, {\it Quant. Inf. Proc.}, in
press. Eprint cond-mat/0204153.

\bibitem{Dodd:01}
{J.L. Dodd, M.A. Nielsen, M.J. Bremner, R.T. Thew}, Eprint quant-ph/0106064.

\bibitem{Bennett:93}
{C.H. Bennett {\it et al.}}, {\it Phys. Rev. Lett.\/} {\bf 70}, 1895
  (1993).

\end{references}
\end{document}